\documentclass[prl,twocolumn,twoside,showpacs,superscriptaddress]{revtex4}
\usepackage{graphicx,amssymb,amsmath,epsf,bm}
\epsfclipon

\newcommand{\p}{\partial}

\begin{document}

\title{Hyperbolicity and the effective dimension of spatially-extended
dissipative systems}

\author{Hong-liu Yang}
\affiliation{%
Institute of Physics,
Chemnitz University of Technology, D-09107 Chemnitz, Germany 
}
\author{Kazumasa A. Takeuchi}
\affiliation{CEA -- Service de Physique de l'\'Etat Condens\'e,~CEN~Saclay,~91191~Gif-sur-Yvette,~France}%
\affiliation{
Department of Physics, The University of Tokyo, 7-3-1 Hongo, Tokyo 113-0033, Japan}%
\author{Francesco Ginelli}
\affiliation{CEA -- Service de Physique de l'\'Etat Condens\'e,~CEN~Saclay,~91191~Gif-sur-Yvette,~France}%
\affiliation{Institut des Syst\`emes Complexes de Paris Ile-de-France, 57-59 Rue Lhomond, 75005 Paris, France}%
\author{Hugues Chat\'e}
\affiliation{CEA -- Service de Physique de l'\'Etat Condens\'e,~CEN~Saclay,~91191~Gif-sur-Yvette,~France}%
\author{G\"unter Radons}%
\affiliation{%
Institute of Physics,
Chemnitz University of Technology, D-09107 Chemnitz, Germany 
}

\date{\today}

\begin{abstract}
Using covariant Lyapunov vectors, we reveal a split of the tangent space
of standard models of one-dimensional dissipative spatiotemporal chaos:
a finite extensive set of $N$ dynamically entangled 
vectors with frequent common tangencies describes all the physically relevant dynamics and is
hyperbolically separated from possibly infinitely many isolated modes
representing trivial, exponentially decaying perturbations. We argue
that $N$ can be interpreted as the number of effective degrees of
freedom, which has to be taken into account in numerical integration and
control issues.
\end{abstract}


\maketitle

Nonlinear dissipative partial differential equations (PDEs) are ubiquitous
in the description of pattern-forming, chaotic, and turbulent 
systems \cite{Cross_Hohenberg-RMP1993}. 
Even though they are formally infinite-dimensional dynamical systems, 
it is now well accepted that their chaotic solutions evolve in an effective
manifold of finite dimension.
For many generic PDEs such as the Kuramoto-Sivashinsky (KS) and
 the complex Ginzburg-Landau (CGL),
it is in fact proven that trajectories are first exponentially attracted
 to a finite-dimensional invariant manifold called the inertial manifold
\cite{InertialManifold}.
This object, however, remains largely formal, as there does not exist
a constructive way of determining of which modes it is composed.
Similarly, trajectories eventually fall into a global attractor 
of finite Hausdorff dimension. For large systems,
this dimension, as well as other quantities measuring the amount of 
chaos in the system, can be estimated via the calculation of Lyapunov exponents.
These dimensions remain, however, global quantifiers. 
One approach to determine which modes actually compose and contribute
to the dynamics is that pursued, e.g., by Cvitanovic {\it et al.}
 \cite{UPOExpansion},
but it is difficult and limited to rather small systems.

A related difficulty lies in
the numerical integration of dissipative PDEs (which remains the primary
way of studying their often chaotic solutions). If the finite-dimensionality
of their attractors justifies that a
numerical study is possible at all, there is no a priori criterion to define
the minimal resolution for a faithful simulation,
and in practice, one typically checks the convergence of results
upon increasing the resolution.

In this Letter, we use covariant Lyapunov vectors (CLVs), recently
made numerically accessible thanks to an efficient
algorithm \cite{Ginelli_etal-PRL2007}, to show that the tangent dynamics
of large KS and CGL systems is essentially characterized by a well-defined set of 
``physical'' modes. Because the covariant vectors span
the intrinsic (Oseledec) subspaces corresponding to each Lyapunov exponent,
and thus allow access to hyperbolicity properties, we are able to show that the physical
modes are decoupled from the remaining set of hyperbolically ``isolated''
degrees of freedom.
In the
context of dissipative partial differential equations, our results imply
that a faithful numerical integration needs to incorporate at least as
many degrees of freedom as the number of such physical modes and that
further increasing the resolution 
increases the number of degrees of freedom associated
with the trivially decaying isolated modes.

\begin{figure}[t]
 \begin{center}
  \includegraphics[clip,width=8.2cm]{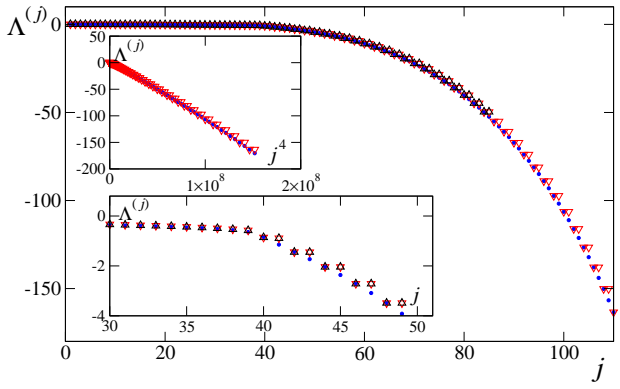}
  \caption{(color online) 
Lyapunov spectrum $\Lambda^{(j)}$ made of exponents arranged by decreasing
value for the one-dimensional KS equation with $L = 96$.
Black upward triangles: $k_{\rm cut} = 42 \cdot 2\pi/L$ and PBC. 
Red downward triangles: $k_{\rm cut} = 85 \cdot 2\pi/L$ and PBC.
Blue dots: $k_{\rm cut} = 170 \cdot \pi/L$ and RBC ($u(L,t) = u(0,t) = 0$).
In all three cases the spectrum yields a positive region with maximum Lyapunov 
exponent $\Lambda^{(1)} \approx 0.09$. 
Top inset: $\Lambda^{(j)}$ vs $j^4$ for the large-$j$ (isolated) modes.
Bottom inset:  close-up around the threshold.}
  \label{fig:KSLyap}
 \end{center}
\end{figure}

We first focus on the one-dimensional KS equation,
taken here as a prototypical dissipative PDE showing space-time chaos
 \cite{Cross_Hohenberg-RMP1993, Chate-ScholarPedia}.
It governs a real field $u(x,t)$ according to
\begin{equation}
 \p_tu = -\p_x^2 u - \p_x^4 u - u \p_x u, ~~~~x \in [0,L] .
\label{eq:KS}
\end{equation}
Figure \ref{fig:KSLyap} shows the Lyapunov spectrum
 for a fixed system size $L = 96$
 but different spatial resolutions and periodic or rigid boundary conditions
(PBC or RBC) \cite{Algorithm}.
The spectrum consists of two parts:
first a smooth region of positive,
 zero, and some negative exponents, then a rather steep
region of negative exponents arranged in steps of two for PBC.
The two regions are separated by an abrupt change in slope
(bottom inset).
Remarkably, the spectra for different spatial resolutions
 overlap with the extra exponents coming from the higher resolution
 simply accumulating in the second region, at the negative end of the spectrum
 (upward and downward triangles).
Thus the threshold index separating the two regions
stays unchanged (here around $j=40$) upon increasing resolution.
Note also that the boundary conditions only change the multiplicity of
 modes in the second region, where every other mode is exactly 
 the same for both PBC and RBC.

\begin{figure}[t]
 \begin{center}
  \includegraphics[clip,width=8.0cm]{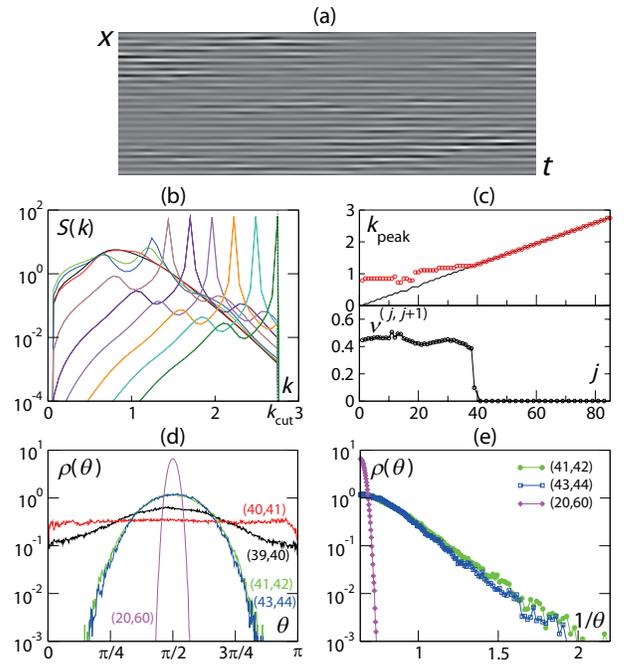}
  \caption{(color online). 
Properties of CLVs for the KS system
($L = 96$, $k_{\rm cut} = 42 \cdot 2\pi/L$, and PBC). 
(a) Spatiotemporal plot of a typical vector in the isolated region
($j=46$, total time $100$). 
(b) Spatial power spectra of vectors of indices $j = 1$, $16$, $32$, $38$, $44$, $52$, $60$, $68$, $76$, $84$ (from left to right in peak position). 
(c) Top panel: peak wavenumber in the power spectra (red circles) 
and $k = [j/2] \cdot 2\pi/L$ (black line). 
Bottom panel: DOS violation fraction $\nu^{(j,j+1)}_\tau$ for pairs of 
neighboring vectors (pairs within the same step are omitted). 
(d) Angle distributions between pairs of vectors.
(e) Same as (d) but with a different abscissa.}
  \label{fig:KSVarious}
 \end{center}
\end{figure}

The above observations suggest that modes in the second region, hereafter called ``isolated'',
for reasons given below, are residual,
highly damped degrees of freedom not necessary to describe properly 
the essential dynamics. In contrast, the modes in the first region, which should be intimately
associated with phase space dynamics, will be called ``physical''. 
In the following, we substantiate this intuition on a rigorous basis
studying the CLVs associated with Lyapunov exponents.  
%
%
The CLVs for the isolated modes, contrary to those
of the physical modes,
possess an approximately sinusoidal, delocalized structure
 as can be seen from their power spectra (Fig.\ \ref{fig:KSVarious}b),
 as well as directly
 from their rather uneventful 
spatiotemporal evolution (Fig.\ \ref{fig:KSVarious}a).
With PBC, the two modes forming a step show the same dominant wavenumber
 with an arbitrary phase shift.
This is not the case of RBC
 where the phase of the sinusoidal structure is fixed at the boundary.
The peak wavenumber $k_{\rm peak}^{(j)}$, which
 linearly increases with $j$ (top panel of Fig.\ \ref{fig:KSVarious}c),
is in fact just the $j$-th wavenumber
 allowed in the given spatial geometry  (multiplicity taken into account):
 $k_{\rm peak}^{(j)} = [j/2] \cdot 2\pi/L$.
For large-enough $j$, 
$\Lambda^{(j)} \sim -(k_{\rm peak}^{(j)})^4 \sim -j^4$ 
(top inset of Fig.~\ref{fig:KSLyap}),
which indicates that the values of the Lyapunov exponents
 of the isolated modes are governed
 by the stabilizing linear term of the KS equation
(i.e. the fourth-order derivative).

The sinusoidal structure of the isolated modes indicates
 that they are nearly orthogonal to each other.
Indeed, distributions of the angle $\theta$
 between pairs of CLVs of indices $j \geq 42$ 
are peaked at $\pi/2$
 and drop rapidly near $0$ and $\pi$.
This is also true if only one of the vectors is taken in the region
 $j \leq 41$, but for any pair of vectors taken from this region,
the angle distribution spans the whole $[0,\pi]$ interval
(Fig.\ \ref{fig:KSVarious}d). 
A careful analysis of the angle distributions reveals that
 those involving isolated modes seem to have an essential singularity
near $0$ (and $\pi$): $\rho(\theta) \sim \exp(-\text{const.}/\theta)$
(Fig.\ \ref{fig:KSVarious}e). Given the sharpness of this behavior,
we cannot conclude about the possibility for these distributions to be 
strictly bounded away from zero.
The above results determine accurately the threshold at $j = 41$, 
and therefore the number of physical modes,
and give the precise definition of 
 physical and isolated modes:
 contrary to the physical modes, 
 the isolated modes do not have any tangencies
 with other isolated modes
 (except the partner in the same step) nor with physical modes.
They can be said to be ``hyperbolically isolated.''
%

The absence of tangency for isolated modes can be confirmed
 from another viewpoint, that of the so-called
 domination of the Oseledec splitting (DOS) \cite{DOS},
 which quantifies, loosely speaking, the degree of dynamical isolation
 of the Oseledec subspaces from each other due to the strict ordering
 of Lyapunov exponents.
Let $\lambda_\tau^{(j)}(t)$ be the finite-time Lyapunov exponent
 averaged over a period $\tau$ around time $t$.
The splitting of the space formed by the vectors
 associated with the modes $j_1$ and $j_2 (> j_1)$
 is said to be dominated
 if $\lambda_\tau^{(j_1)}(t) > \lambda_\tau^{(j_2)}(t)$ holds
 \textit{for all} $t$ with $\tau$ larger than some finite $\tau_0$.
It is mathematically proven that DOS implies absence of tangency
 between the Oseledec subspaces, or the CLVs \cite{DOS}.
To quantify DOS, we define, following \cite{Yang_Radons-PRL2008}
$
\Delta\lambda_\tau^{(j_1,j_2)}(t) \equiv \lambda_\tau^{(j_{\rm min})}(t) 
- \lambda_\tau^{(j_{\rm max})}(t)$
with $j_{\rm max} \equiv \max (j_1,j_2)$ and $j_{\rm min} \equiv \min (j_1,j_2)$
and measure the time fraction of DOS violation
$\nu^{(j_1,j_2)}_\tau = \langle \Theta( \Delta\lambda_\tau^{(j_1,j_2)}(t) ) \rangle$,
 where $\Theta(z)$ is the step function
 and $\langle\cdots\rangle$ denotes the time average.
The result is shown in the bottom panel of Fig.\ \ref{fig:KSVarious}c for pairs of
neighboring exponents:
$\nu^{(j,j+1)}_\tau$ with $\tau = 0.2$ drops sharply near the threshold
 and becomes \textit{strictly} zero for $j \geq 41$.
In fact, $\nu^{(j,j')}_\tau$ stays zero for any pairs with $j$ or $j' > 41$,
 though in some cases slightly larger values of $\tau$ are required.
This confirms the absence of tangencies of the isolated modes,
already seen from the angle distributions, and the number of physical
modes.

We now turn our attention to a different case in order to test the 
validity of our results beyond the simple KS equation.
Let us consider the CGL equation, whose universal relevance and genericity 
is now well-established
 \cite{Cross_Hohenberg-RMP1993,Aranson_Kramer-RMP2002}. In one space dimension,
it governs a complex field $W(x,t)$ according to:
\begin{equation}
 \p_tW = W - (1+i \beta)|W|^2W + (1+i \alpha)\p_x^2 W \;.
\label{eq:CGL}
\end{equation}
In the following 
we consider a so-called ``amplitude turbulence'' regime
 \cite{Shraiman_etal-PhysD1992}, 
i.e. a strongly chaotic
regime where amplitude and phase modes evolve on rather short time- and 
length-scales. (Results for other regimes, such as phase turbulence, 
will be presented elsewhere \cite{TBP}.)
Specifically, we use $\alpha = -2.0, \beta = 3.0, L = 64$, and PBC.

\begin{figure}[t]
 \begin{center}
  \includegraphics[clip]{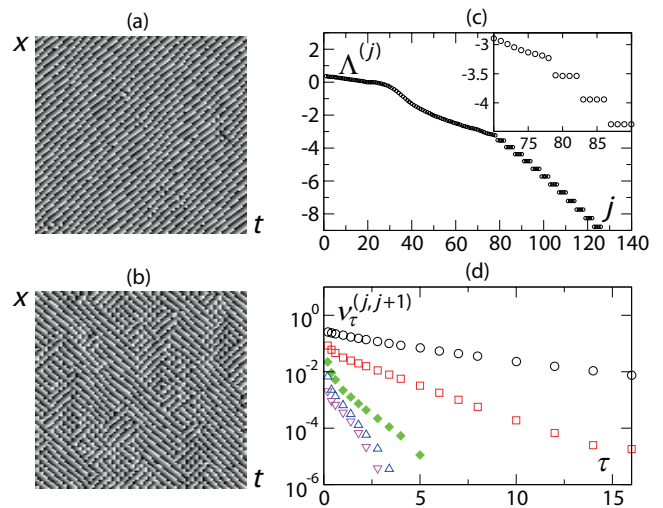}
  \caption{(color online)
CGL equation in the amplitude turbulence regime 
($L = 64$, $k_{\rm cut} = 31 \cdot 2\pi/L$, and PBC). 
(a,b) Spatiotemporal plots of the phase component of a typical vector $j = 91$ in the isolated region
(same trajectory, but at two distant periods of time, during a total time of $20$ for each plot).
(c) Lyapunov spectrum; inset: close-up around threshold.
(d) Time fraction $\nu^{(j,j+1)}_\tau$ of DOS violation, as a function of $\tau$
($j=78$, 82, 86, 90, 94, from top to bottom).}
  \label{fig:CGLVarious}
 \end{center}
\end{figure}

For sufficiently high spatial resolution,
 the Lyapunov spectrum indeed shows an isolated, 
stepwise region as for the KS equation (Fig.\ \ref{fig:CGLVarious}c),
 but here the multiplicity of each step is four and the 
spatiotemporal evolution of isolated modes reveal patches of traveling waves
(Fig.\ \ref{fig:CGLVarious}ab). (The effects of spatial resolution
and/or boundary conditions are also similar to our observations on the KS 
equation.)
As for the KS equation, this is in agreement with 
the linear stability analysis:
normal modes are traveling waves
 of the form $\exp[i(\pm kx-\omega_k t) + \Lambda t]$
 with $\Lambda = 1-k^2$ and $\omega_k = \alpha k^2$.
Indeed, the values of the Lyapunov exponents in the isolated region behave
like this
 and the vectors are composed of traveling waves
propagating at a velocity of $\pm \omega/k$. Compared to the KS case,
the additional multiplicity of two comes from the degeneracy
 between $k$ and $-k$ modes.
Moreover, this degeneracy implies that
 these two modes are in fact mixed up in a single vector:
 isolated vectors are either in the pure $k$ mode
(Fig.\ \ref{fig:CGLVarious}a),
 in the pure $-k$ mode, or patches of the two
(Fig.\ \ref{fig:CGLVarious}b).

In spite of these differences, 
the isolated modes for CGL remain dynamically isolated
 from any other mode.
Although $\nu^{(j,j+1)}_\tau$ measured with small $\tau$ is not zero
 in the isolated region,
 a clear threshold is found when considering larger $\tau$ values:
Contrary to what happens for $j_1,j_2 \leq 86$, all $\nu^{(j_1,j_2)}_\tau$
 for $j_2 \geq 87$ show a decrease faster than exponential,
 indicating the existence of a finite $\tau$ beyond which
 $\nu^{(j_1,j_2)}_\tau$ is zero (Fig.\ \ref{fig:CGLVarious}d).
Consequently, we can also define the isolated modes by their DOS 
as for the KS equation, and determine exactly the number of physical modes
(here at $j=86$). Indeed, physical modes have 
no tangencies with any isolated modes ($j\ge 87$). Moreover,
 angle distributions involving isolated modes show, like for the KS equation,
the essential singularity $\sim \exp(-\text{const.}/\theta)$ near tangency.

We now discuss our results.
We first note that the threshold separating physical from isolated
modes does {\it not} coincide with the appearance of steps
in the Lyapunov spectrum (for PBC). 
Lyapunov exponents alone can only provide a good guess:
For the KS and CGL systems treated above,
the first steps start respectively at $j=40$ and $j=79$, whereas the 
exact thresholds are at  $j=41$ and $j=86$. Indeed, 
a closer scrutiny of the exponents reveal that the first steps are 
actually not perfect.

\begin{figure}[t]
 \begin{center}
  \includegraphics[clip,width=8.6cm]{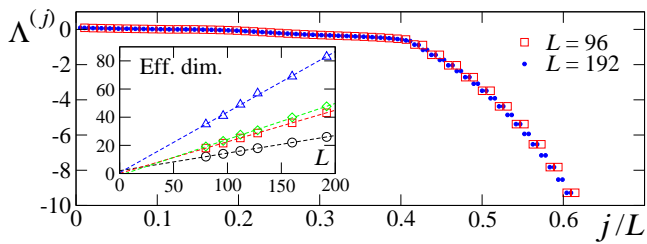}
  \caption{(color online). 
Extensivity of the Lyapunov spectrum for the KS equation with PBC. 
Inset: quantities indicating effective dimensions of the system: number of non-negative exponents (circles), 
Kaplan-Yorke dimension (squares), 
metric entropy (diamonds, multiplied by $50$), 
and number of physical modes (triangles).}
  \label{fig:KSExtensivity}
 \end{center}
\end{figure}

Let us now specify the implication of the lack of tangencies
 for the isolated modes.
Suppose that we add to the dynamics an infinitesimal perturbation 
along the CLVs of some isolated modes.
Then this perturbation decays exponentially to zero as indicated
 by their negative Lyapunov exponents,
 and the absence of tangencies implies that
 this does not induce any perturbation along directions spanned
 by the other Lyapunov modes.
In contrast, perturbations along physical modes will propagate to other physical modes
 through tangencies between them, and could eventually induce activity in the
 modes associated with positive exponents, growing to considerably affect phase space dynamics 
 even if the initial perturbation was made in the direction associated with negative exponents.
In this sense, the dynamics corresponding to the physical modes
 is highly entangled, but completely decoupled from
 the decaying dynamics of the isolated modes.
Therefore, all the degrees of freedom associated with physical modes are
 necessary to faithfully describe phase space dynamics, while adding further more degrees of
 freedom associated with isolated modes 
 does not affect phase space dynamics in any significant way.
The number of physical modes
 scales linearly with the system size $L$: rescaled Lyapunov spectra
 collapse both in the physical and in the isolated region with the 
 stepwise structure retained (Fig.~\ref{fig:KSExtensivity}). 
In particular, the dimension density defined from the number of physical modes is 
 larger than others, for instance being almost twice the Kaplan-Yorke dimension \cite{KY}
 (inset of Fig.~\ref{fig:KSExtensivity}). It is therefore natural to interpret the number of
 physical modes as an embedding dimension of the global attractor. 
Furthermore, we speculate, pending mathematical rigor, that the number of physical modes could be related 
 to the dimension of the inertial manifold, which is a positively invariant and
 exponentially attracting smooth manifold embedding the global attractor
(the subspace spanned by the physical modes would then be the local linear 
approximation of the inertial manifold \cite{IM2}).
While this conjecture would drastically lower existing estimates
 for the dimension of the KS inertial manifold
 $D \leq \text{const.} \times L^{2.46}$ \cite{IM2},
 it is however in line with
 the extensivity of chaos already observed in the past
 for the same models \cite{Manneville-LNP1985, Egolf}. 

Our results suggest that a faithful numerical integration of PDEs needs to incorporate at least  
the degrees of freedom associated with all physical modes. 
Moreover, the extensivity of the number of physical modes implies that
the minimal resolution evaluated in systems of moderate size
carries over to arbitrarily large system size $L$.
Our obtained number of effective degrees of freedom would also be helpful to determine
the minimal number of constraints 
necessary for a full control of a continuum system, with applications to real situations like e.g. the 
suppression of ventricular fibrillation.
%

In summary, we have shown, using Lyapunov analysis, 
that the tangent space of two representative nonlinear dissipative PDEs 
systems can be divided into two parts:
 a finite dimensional manifold spanned by strongly interacting physical modes,
and the remaining set of isolated, strongly damped modes. We demonstrated
that isolated modes are hyperbolically separated from all other modes,
 and thus satisfy
the property of domination of Oseledec splitting.
Similar results were obtained
 also for a chain of diffusively-coupled tent maps (not shown, \cite{TBP}). 
We have interpreted the number of physical modes as an
 embedding dimension of the global attractor.
The extensivity of this dimension could also be of interest in view of 
the studies by Egolf {\it et al.}
 \cite{Egolf} arguing about the ``building blocks'' of spatiotemporal chaos.
We hope our results will trigger work to clarify these issues at the mathematical level,
as much as numerical investigations of other dissipative systems,
like those yielding fully developed turbulence, for which no rigorous proofs are known about
the existence of an inertial manifold.
Besides their theoretical importance,
our results are also useful for the numerical integration and control issues of dissipative PDEs.


\begin{thebibliography}{99}
\bibitem{Algorithm}
A spectral method with discrete Fourier or sine transform
 up to a cutoff wavenumber $k_{\rm cut}$ is used in order to avoid aliasing.
Time is integrated using the operator-splitting method with a time step $0.005$.
Data are typically recorded over a period of $\approx 10^5$
 after a transient of $\approx 5 \times 10^3 \times k_{\rm cut}$.
\end{thebibliography}
\end{document}